# Non-proportional scintillation response of NaI:Tl to low energy X-ray photons and electrons

I.V. Khodyuk[1,2], P.A. Rodnyi[2], and P. Dorenbos[1]
[1]Faculty of Applied Sciences, Delft University of Technology, 2629JB Delft, The Netherlands
[2]Saint Petersburg State Polytechnical University, St. Petersburg, 195251 Russia

Non-proportional response of the scintillation yield of NaI:Tl was measured using highly monochromatic synchrotron irradiation ranging from 9 to 100 keV. Special attention is paid to the X-ray escape peaks. They provide us additional information about non-proportional response in the range 0.9 to 12 keV. A rapid variation of the non-proportional response curve is observed near the Iodine K-electron binding energy. A dense sampling of data is performed around this energy and that data are used to apply a method, which we call K-dip spectroscopy. This method allows us to derive the electron response curve of NaI:Tl down to energies as low as 30 eV. A comparison of our data with data of others employing different methods is made. Advantages, limitations and peculiarities of presented techniques and methods are discussed.

**I. Introduction**

NaI:Tl inorganic scintillation crystals were discovered in 1948 [1] and are still today the best known and most widely used scintillators. Despite the large number of scintillating compounds only few of them [2, 3] can compete with NaI:Tl in terms of light output and energy resolution. Large amount of research has been done to unravel and to understand the scintillation mechanism in NaI:Tl, but many aspects are still not fully understood. For example, more then 50 years ago [4] it was found that the amount of photons emitted in the scintillation spark caused by absorption of an X-ray, a γ-quantum, or a particle in NaI:Tl is not precisely proportional to its energy. This finding appears important because it causes the energy resolution achievable with scintillation material to be worse than what might be expected on purely statistical grounds [5]. Although the phenomenon of non-proportional response (nPR) and its relation with energy resolution (R) has been studied quite intensively [6-14] there are still many major gaps in our understanding of the underlying physics. Accurate data from dedicated experimental techniques are needed to reveal the true origin of nPR and energy losses inside the solid state. We aim to develop models on non-proportionality that may help us in improving the scintillation properties of existing materials and that helps us in our search for new highly effective and low energy resolution scintillators.

Since gamma radiation produce fast electrons in the solid state, nPR as function of gamma energy is a direct consequence of the more fundamental nPR as a function of primary electron energy. A method to study the electron response of a scintillator is the Compton Coincidence Technique (CCT) introduced by Valentine and Rooney [15] and further developed by Choong *et al.* [16]. In a Compton scattering event the scattered gamma ray escapes the scintillator and the photon yield produced by the Compton electron alone is determined with the CCT as function of its energy. The main advantages of this method are the wide Compton electron energy range, usually from 3 to 450 keV, that is covered, and that the results are not affected by the surface of the scintillator. However, using CCT, it is not possible to obtain reliable data on the electron response at energies below 3 keV.

In this work we will demonstrate that by measuring the photon-nPR of the scintillator using highly monochromatic synchrotron X-rays, it is possible to obtain electron-nPR data starting from energy as low as 30 eV without a disturbing influence of the scintillator surface. Accurate experimental data is especially important in this low energy range because there the most dramatic change in scintillator efficiency and nPR is expected. We are not aware of any other experimental method that provides information on electron response down to that low energy.

We will start from a description of the experimental setup used to obtain data. The geometry of the sample and how it is packed will be described. In this work we will define and introduce different types of photon-nPR curves. The photon-nPR curve obtained using direct observation of photopeaks from total absorption of highly monochromatic X-ray synchrotron irradiation will be presented. We will call this the photopeak-nPR curve. Special attention is paid to the escape peaks and how to use them to get additional information about photon-nPR in the low energy range. So-called escape-nPR curves will be shown. The method to estimate electron-nPR analogous to the one used by Collinson and Hill [17] and later by Wayne et al. [18], which we called K-dip spectroscopy [19], is described in detail and used to reconstruct the so-called K- electron-nPR curve of NaI:Tl down to electron energies as low as 30 eV. A comparison of our data with the data of other authors is presented. Advantages, limitations and peculiarities of our techniques will be discussed. The aim of this work is to provide new data and methods to obtain those. It is not our aim to provide a complete explanation of the observed nPR curves.

**II. Experimental methods**

NaI:Tl is hygroscopic and to study its photon-nPR down to X-ray energies of 9 keV an X-ray assembly was manufactured by the company Saint-Gobain Crystals&Detectors. Since we intended also to exploit X-ray escape peaks for our studies, a small 10 mm diameter and 2 mm thick NaI:Tl crystal was used to increase the probability of X-ray fluorescence escape. As entrance window for the X-rays, 220 μm thick Beryllium was used in order to avoid too much absorption at low energy X-ray irradiation. The crystal is sealed in an Al housing with 1 mm thick quartz window and its 2 mm edges were covered with a white reflector to maximize the photon collection at the photomultiplier tube (PMT) photocathode.

The number of photoelectrons $N_{phe}^{PMT}$ per MeV of absorbed energy produced in a Hamamatsu R6231-100 PMT by NaI:Tl was determined by comparing the position of the $^{137}$Cs 662 keV photopeak or of the $^{241}$Am 59.5 keV photopeak in recorded pulse height spectra with the mean value of the so-called single photoelectron pulse height spectrum. The procedure has been described in detail by de Haas et al. [20]. To collect as much of the emitted light as possible, the NaI:Tl scintillator was optically coupled to the entrance window of the PMT and the shaping time of an Ortec 672 spectroscopic amplifier was set at 10 μs.

To measure the pulse height spectra at many finely spaced energy values between 9 keV and 100 keV, experiments at the X-1 beamline at the Hamburger Synhrotronstrahlungslabor (HASYLAB) synchrotron radiation facility in Hamburg, Germany were carried out. The scheme of the experimental set-up is presented in Fig. 1. A highly monochromatic pencil X-ray beam in the energy range 9 – 100 keV was used as excitation source. A tunable double Bragg reflection monochromator using a Si[511] and Si[311] set of silicon crystals providing an X-ray resolution of 1 eV at 9 keV rising to 20 eV at 100 keV was used to select the X-ray energies. The beam spot size was set by a pair of precision stepper-driven slits, positioned immediately in front of the sample coupled to the PMT. For all measurements, a slit size of 50 × 50 μm$^2$ was used. The PMT was mounted on an X-Y table capable of positioning with a precision of <1 μm in each direction. Prior to each measurement, the position of the PMT was adjusted to achieve as high count rate as possible. The intensity of the synchrotron beam was reduced in order to avoid pulse pileup. A lead shielding was used to protect the sample from receiving background irradiation which otherwise appeared as a broad background in our pulse height spectra.

To record synchrotron X-ray pulse height spectra of NaI:Tl, a Hamamatsu R6231-100 PMT connected to a homemade preamplifier, an Ortec 672 spectroscopic amplifier and an Amptek 8000A multichannel analyzer (MCA) were used. The quartz window of the NaI:Tl assembly was optically coupled to the window of the PMT with Viscasil 600 000 cSt from General Electric. The NaI:Tl assembly plus PMT entrance window was covered with several layers of ultraviolet reflecting Teflon tape (PFTE tape) forming an "umbrella" configuration [21]. Scintillation photons reflected from the photocathode are then reflected back by the umbrella thus enhancing detection efficiency. All measurements were carried out at room temperature and repeated several times.

Corrections were made for channel offsets in the pulse height measurement. The offset was measured by an Ortec 419 precision pulse generator with variable pulse height attenuation settings.

### III. Results and discussion

A. Data analysis

In Fig. 2 a typical pulse height spectrum recorded with NaI:Tl at 40 keV monochromatic X-ray irradiation is shown. The photopeak labeled "a" is fitted with a single Gaussian shaped curve from which the position of the maximum of the peak and its full width at half maximum (FWHM) intensity is obtained. This type of pulse height spectra was recorded for a large set of X-ray energies between 9 keV and 100 keV providing data on scintillation photon yield, from which the photon-nPR can be obtained, and data on scintillator resolution.

To get additional information of the photon-nPR at low X-ray energies, escape peaks "b" in Fig. 2 were analyzed. X-ray photons of energy between the Iodine K-electron binding energy $E_{KI}$=33.169 keV and 100 keV interact with matter almost exclusively by means of the photoelectric effect. After interaction the electron is ejected from the atom's K-shell, leaving a hole. As the atom returns to its stable lowest energy state, an electron from one of its outer shells jump to the hole in the K-shell, and in the process giving off a characteristic X-ray photon or Auger electrons. In the case that characteristic X-ray photons escape the bulk of the crystal we observe an escape peak like the peaks labeled "b" in Fig. 2. The ranges of Auger electrons are too short to escape the bulk of the material and we do not consider Auger electron escape here.

The deposited energy $E_d$ associated with events counted in the escape peak is then the energy of the X-ray photon $E_X$ from the synchrotron minus the energy $E_e$ of the escaped X-ray:

$$E_d = E_X - E_e. \quad (1)$$

In fitting escape peaks we assumed 5 possible fluorescent transitions in Iodine. The scheme of Fig. 3 illustrates the $K_{\alpha 1}$, $K_{\alpha 2}$, $K_{\beta 1}$, $K_{\beta 2}$ and $K_{\beta 3}$ transitions between the shells of an Iodine atom. The energies and probabilities for the transitions used in the fitting of the escape peaks are listed in Table 1. Assuming that every type of escape results in a Gaussian shaped escape peak, we used a sum of five Gaussian peaks to fit the observed escape peaks. The widths of the five Gaussian peaks were assumed all the same like in [22]. Result of the fitting is shown in the inset of Fig. 2. After fitting, the experimental values of the maxima are known for the $K_{\alpha 1}$ $K_{\alpha 2}$ and $K_{\beta 1}$, $K_{\beta 2}$ and $K_{\beta 3}$ escape peaks.

For this work we are interested in the weighted mean position of the two $K_\alpha$ peaks and the three $K_\beta$ peaks. For NaI:Tl it turns out that the position of the resulting $K_\alpha$ and $K_\beta$ maxima are located close to the values estimated by direct fitting of the two escape peaks with two Gaussians. Usage of two Gaussians instead of five would have simplified our fitting procedure, but for other scintillators $K_\alpha$ and $K_\beta$ escape peaks are not so well separated as in Fig. 2 and then fitting with five Gaussians is the preferred method.

Now we need to know what $E_d$ energies correspond to the found $K_\alpha$ and $K_\beta$ maxima. Based on the energies and probabilities from Table 1 the mean values of the escape energies $E_{K_\alpha}$ and $E_{K_\beta}$ were calculated, and then Eq. (1) provides $E_d$. Repeating the same procedure for all $E_X$ above the $E_{KI}$ we obtain the photon yield curve as function of $E_d$.

B. Photopeak non-proportional response

The number of photoelectrons $N_{phe}^{PMT}$, created in the PMT using synchrotron X-rays was determined at energies between 9 and 100 keV with a 5 keV step size. In the energy range 9 to 12 keV, a 1 keV step size was used. A much finer step size of 25 eV was used around $E_{KI}$ =33.169 keV, because interesting features are observed around that energy. Figure 4 shows $N_{phe}^{PMT}$, created in the PMT as derived from the photopeak position in the pulse height spectra versus $E_X$. With this method of plotting data, the $N_{phe}^{PMT}$ appears to increase proportionally with $E_X$. In the inset of Fig. 4, the data near $E_{KI}$ has been plotted on an expanded scale. Now, a clear step can be seen in the $N_{phe}^{PMT}$ exactly at $E_{KI}$.

In this work we define the photopeak-nPR of NaI(Tl) at $E_X$ as the $N_{phe}^{PMT}$ /MeV observed at energy $E_X$ divided by the $N_{phe}^{PMT}$ /MeV observed at $E_X$ = 662 keV energy. The nPR will be expressed as a percentage value. Figure 5 shows the thus obtained photopeak-nPR curve as a function of $E_X$. Figure 6 shows the same photopeak-nPR curve but with a dense sampling at energies around $E_{KI}$. A clear dip is observed that we name the K-dip. As will be shown further in this paper we can derive valuable data on the electron response curve down to energies as low as 30 eV from a detailed analysis of the photopeak-nPR around such K-dip. We have named such analysis K-dip spectroscopy [19].

The shape of the photopeak-nPR curve is similar to results reported before [8], i.e., a linear increase from 111.2% to 115.8% with decrease of $E_X$ from 100 keV to 50 keV followed by a drop in the range 30 – 45 keV with a local minimum of 114.1% at 34.5 keV. Next the photopeak-nPR increases up to 117.2% at 20 keV followed by a steep decrease of the response with further decrease of $E_X$. The nPR at 9 keV is 111.5% which is almost equal to the nPR at 100 keV. So there appears a drop of 5.7% in the photopeak-nPR is going from 20 keV to 9 keV which is of interest for further investigation.

The energy resolution $R(E_X)$ of the X-ray photopeaks is plotted in Fig. 7 versus $E_X$. Starting from 9 keV to 100 keV $R$ decreases from 21.9% to 6.7%. A clear step-like change of almost 0.2% can be seen at $E_X$ around $E_{KI}$. $R(59.5 keV)$ measured using the $^{241}$Am source, was 10.1 %. With 59.5 keV synchrotron X-ray irradiation a value of 8.1% is observed. We attribute the improvement to the fact that the synchrotron X-rays are collimated but the $^{241}$Am gamma rays are absorbed throughout the bulk of the crystal. In the latter case inhomogeneities in the crystal properties or light collection properties provide an additional contribution to the energy resolution.

In Fig. 8 the same data as in Fig. 7 are shown but now displayed versus $N_{phe}^{PMT}$. The solid curve represents the theoretical limiting resolution due to the always present Poisson statistics in the number of detected photons [8]:

$$R_{stat} = 2.355 \sqrt{\frac{1+\nu}{N_{phe}^{PMT}}}, \qquad (2)$$

where $\nu = 0.25$ is the contribution from the variance in the gain of the Hamamatsu R6231-100 PMT.

Figure 8 shows that the $R(E_X)$ just below $E_{KI}$ is quite close to the theoretical limit. Besides the contribution from $R_{stat}$ there are other contributions to $R$ [8]. The contribution from inhomogeneity in the scintillator light yield and light collection is regarded negligible because of the collimated X-ray beam geometry. What remains is an intrinsic contribution $R_{np}$ due to the nPR of NaI:Tl. This contribution can be calculated with:

$$R_{np} = \sqrt{R^2 - R_{stat}^2} \ . \qquad (3)$$

C. Escape non-proportional response

So far we only used information from the photopeaks in pulse height spectra as function of $E_X$ to obtain the photon-nPR curve. One may also use information derived from the escape peaks to obtain a photon-NPR curve. Using Eq. (1) and the procedure described after Eq. (1), we can construct a, what we call, an escape-nPR curve for NaI:Tl from the $K_\alpha$ and $K_\beta$ escape peaks data as a function of $E_d$. The results are shown in Figs. 9 and 10. The $K_\alpha$ escape-nPR data from the $K_\alpha$ escape peak position analysis as function of $E_d$ match the data obtained from the photopeaks analysis as function of $E_X$ well in the energy interval 9 to 12 keV, as can be seen in Fig. 9. We explain this as follows. At $E_X$ below $E_{KI}$ = 33.169 keV, the by far most probable interaction of the incident X-ray with NaI is the photoelectric absorption by an L-shell electron of Iodine. The interaction creates a photoelectron with energy $E_{phe}^L$ equal to:

$$E_{phe}^L = E_X - E_{LI}^i , \qquad (4)$$

where $E_{LI}^i$ is a binding energy of one of the three L-subshells of Iodine indicated by the superscript $i$. This photoelectron produces an amount of light given by $L_{phe}^L$. The interaction also creates a hole in the L-shell which initiates a cascade of secondary processes involving the emission of Auger electrons and possibly low energy X-rays. In any case the total energy of the hole will be dissipated in the scintillator and converted to an amount of light given by $L_{cascade}^L$.

In the case of $K_\alpha$ X-ray escape we have photoelectric interaction at the Iodine K-shell. The energy of the photoelectron $E_{phe}^K$ in this case will be:

$$E_{phe}^K = E_X - E_{KI} . \qquad (5)$$

Next, a transition occurs of an electron from the L-shell to the K-shell with emission of the $K_{\alpha 1}$ or $K_{\alpha 2}$ X-ray which escapes the scintillator. Again a hole is created in the L-shell which produces as above the same amount $L_{cascade}^L$, of scintillation photons. Therefore, in both cases, i.e., photoelectric absorption at the Iodine L-shell or at the K-shell with subsequent X-ray escape, we have an L-shell photoelectron or a K-shell photoelectron of the same energy producing in first approximation the same amount of photons $L_{phe}$ and we have in both cases an L-shell hole producing $L_{cascade}^L$ amount of light. Therefore, as a first approximation both the photopeak-nPR should be about the same as the $K_\alpha$ escape-nPR in the energy range below $E_{KI}$. In second approximation, we can not treat the $K_\alpha$ escape-nPR as completely the same as a photopeak-nPR. $K_\alpha$ X-ray fluorescence is caused by a transition of an electron from the $L_3$ (2p$_{3/2}$ orbital) or the $L_2$ (2p$_{1/2}$ orbital) subshell to the K (1s orbital), see scheme of Fig. 3. The probabilities and the energies for the two transitions listed in Table 1 are not equal. The transition from the $L_1$ (2s orbital) to the K-shell is dipole forbidden and we can ignore that possibility. Because of the difference in the probability of a hole to be created in the $L_1$, $L_2$ or $L_3$-subshell between photopeak-nPR and escape-nPR some deviation can arise.

As can be seen in Figs. 9 and 10 there is a dip in the $K_\alpha$ escape-nPR with minima at energy about 5.5 keV. We call this the L-dip which is analogues to the K-dip that can be seen at energy around $E_{KI}$. The energy for both the K-dip and L-dip correspond with the discontinuities in the attenuation length curve presented in Fig. 9 and with the iodine electron binding energies indicated by arrows in Fig. 10. As compared to the K-dip, the L-dip is not as sharply defined because of the presence of three L-subshells with slightly different binding energies.

Next to the $K_\alpha$ escape-nPR we can define another photon-nPR that is based on $K_\beta$ escape peak analysis. We call this the $K_\beta$ escape-nPR, and the results are shown in Figs. 9 and 10. The $K_\beta$ escape-nPR data do not overlap the $K_\alpha$ escape-nPR but the two data sets cross each other around the Iodine $L_3$-subshell energy of 4.557 keV, shown as arrow $L_3$ in Fig.10. Above we have argued that the $K_\alpha$ escape-nPR at energies below $E_{KI}$ in first approximation is similar to the photopeak-nPR. For the same reasoning the $K_\beta$ escape-nPR at energies below the energy of the $L_3$ subshell is as a first approximation the same as the photopeak-nPR. It is therefore not a coincidence that the $K_\beta$ escape-nPR crosses the $K_\alpha$ escape-nPR near the $L_3$ subshell energy. We anticipate similar behavior for other scintillation crystals. That reasoning is now as follows. For the photopeak event at energies $E_X$ below the $L_3$ subshell a hole is created in the Iodine M or N shells and the light yield observed is from the cascade of the hole in those shells plus the light produced by the photoelectron from those shells. For the $K_\beta$ escape peak event the light yield is from a K-shell photoelectron plus also from the cascade following the creation of a hole in the M, N subshells. Again in first approximation similar total light yield is expected for the beta escape event and the photopeak event. Analogue to the K and L-dips we can determine an M-dip with minimum at energy about 2 keV. M-dip seems to have a small shift to the higher energies in respect to the X-ray attenuation length in NaI Fig. 9 and M-shell binding energies shown in Fig. 10. One of the reasons for that can be the fact, that in second approximation the shape of the photopeak-nPR can differ from the shape of the $K_\beta$ escape-nPR because of the Sodium K-shell at energy 1.071 keV.

Values for the $K_\beta$ escape-nPR in the energy range above the Iodine $L_3$-subshell energy are higher then the values for the $K_\alpha$ escape-nPR in Figs. 9 and 10. We explain this as follows. For the same value of deposited energy $E_d$ in the case of $K_\beta$ escape the created photoelectron has higher energy as in the case of $K_\alpha$ escape. The difference between the photoelectron energy is equal to the difference between the electron binding energies of L and M or N-subshell electrons. The strong increase of photon-nPR in the range 9-20 keV implies that the scintillation efficiency increases with $E_X$ which suggests an increase with primary electron energy. In other words higher energy electrons are more efficient in producing scintillation light. We believe that this is the reason that the $K_\beta$ escape-nPR curve is running above the $K_\alpha$ escape-nPR curve.

We have now demonstrated that by piecing together the $K_\beta$ escape-nPR below $E_{LI}$, the $K_\alpha$ escape-nPR between $E_{LI}$ and $E_{KI}$ and the photopeak-nPR between 9 keV and 100 keV we obtain the overall photon-nPR from 1 keV to 100 keV which could be further extended by utilizing radioactive sources up to say 10 MeV energy. This overall photon-nPR and the three other types of photon-nPR curves in Figure 9 reveal quite detailed and complex features especially near the binding energies of the iodine subshells. To further understand those features one needs to know the scintillation photon yield of NaI as function of electron energy. Then with Monte Carlo (MC) simulation the distribution of created primary and secondary electrons upon interaction with an X-ray photon in NaI:Tl can be simulated for each of the three photon-nPR, and the total number of scintillation photons calculated. From this one should be able to reproduce each of the three nPR curves. The other way around one could also use the observed nPR curves to deduce the photon yield as function of electron energy [23]. Below we will demonstrate that our data above $E_{KI}$ in Fig.6 enables us to derive the the scintillator response to electrons and then also the electron-nPR curve down to energies as low as 30 eV.

D. K-electron non-proportional response

Measuring the scintillator pulse height at energies below 9 keV is difficult because of the short attenuation length of X-ray absorption, and X-rays either do not transmit the Be-window or at best are absorbed close to the scintillator surface. However, then the surface may affect the scintillation yield. Usage of energies above 9 keV assures us that we are studying the properties of the bulk, not influenced by surface effects [24]. We therefore need other techniques to determine the nPR at energies below 9 keV.

To utilize K-dip spectroscopy, we need precise measurement of the photopeak-nPR in the energy range just above $E_{KI}$, i.e., like the results of the 25 keV step size measurement shown in Fig. 6. The drop of the photopeak-nPR in the range 33.0 – 34.5 keV is more then 1%. Showing error bars would blur all data and are therefore not shown in Fig. 6. In the presented range the average error is less then 0.05%. The main advantage of our method to obtain data compared to that from other methods is the high precision of the results.

The method can be described as follows. An X-ray, photoelectrically absorbed by Iodine, leads to the creation of a number of electrons; a photoelectron plus several Auger electrons. We assume that these electrons then act independently from each other. With this we mean that the number of photons $L_{phe}$ created by the photoelectron is not affected by the presence of the Auger electrons emitted from the same atom and vice versa. The total photon yield is the sum of the photons produced by the complete set of electrons. The response of a scintillator is then equivalent to the sum of two main interaction products: 1) a K-shell photo electron plus 2) the electrons emitted due to the sequence of processes following relaxation of the hole in the K-shell, the so-called K-cascade response. Our strategy is to employ X-ray energies just above $E_{KI}$. The K-cascade response is assumed independent from the original X-ray energy. This response is found by tuning the X-ray energy very close above $E_{KI}$. By subtracting the K-cascade response from the total X-ray response we are left with the response in photoelectrons from the K-shell photo-electron alone with energy $E_X - E_{KI}$. The electron-nPR curve is then obtained by the $N_{phe}^{PMT}$/MeV at the energy of the K-photoelectron divided by the $N_{phe}^{PMT}$/MeV measured at 662 keV.

Figure 11 shows the K-electron-nPR for NaI:Tl using our K-dip spectroscopy method. An alternative method to obtain an electron-nPR curve is by means of CCT. Rooney and Valentine pioneered this method and used it to determine the Compton electron-nPR curve of NaI:Tl [15, 25]. Their results are also shown in Fig. 13. Choong *et al.* [16] further developed the CCT and the improved setup named SLYNCI was used by Hull *et al.* [26] to determine the Compton electron-nPR of different NaI:Tl crystals. Because for different NaI:Tl crystals the nPRs vary, in Fig. 11 we have shown the highest and the lowest values presented in [26]. The data measured by two different groups using the same method but different setups are in a good agreement with our data at energy above 20 keV. Below 20 keV, the Compton electron-nPR curve measured by Rooney and Valentine is at higher value. CCT and SLYNCI do not provide reliable electron-nPR data below 3 keV, and here we think that our data is most reliable.

All presented curves in Fig. 11 have the same appearance. Starting with 100 keV the electron-nPR increase until a maximum is reached at 15 keV and then at even lower energies it decreases again. The increase of the electron-nPR is about 15% for the data reported by Rooney and Valentine in the range 10 – 100 keV; from 10% to 13% for Hull *et al.* in the range 15 – 100 keV; 12% for Wayne *et al.* in the range 15 -100 keV and 13% for K-electron-nPR in the range 10 – 100 keV. With further decreasing of the electron energy, the nPR starts to drop rather fast. In the low energy range, below 10 keV, electron response taken from Wayne et al. [18] is showing lower values as compared with K-dip spectroscopy results. But, considering large error margins for both methods in the electron energy range below 1 keV, we can conclude that out K-dip spectroscopy data are in a good agreement with results from the modified Collinson and Hill method. However, our method provides much more data points with higher accuracy and extending to lower energy.

**IV. Conclusion**

We have measured the non-proportional response (nPR) of NaI:Tl to highly monochromatic X-ray photons in the energy range 9 – 100 keV. By utilizing the photopeak, the $K_\alpha$ escape peak, and the $K_\beta$ escape peaks in pulse height spectra we introduced three different types of strongly related non-proportionality curves. It enables us to obtain a good estimate for the non-proportionality curve of NaI:Tl to X-ray photons down to energies as low as 1 keV. Information that could not be obtained utilizing a 1 keV X-ray source because of unavoidable affects of the scintillator surface. We paid special emphasis to the scintillator response near the K-electron binding energy of the Iodine. From this data, we have inferred the non-proportional response curve (K-electron-nPR) of NaI:Tl to the iodine K-shell photoelectron in the energy range 0.03 – 65 keV. We have named this method K-dip spectroscopy, and it provides us with information on the electron response down to 30 eV. From 65 keV to 10 keV, K-electron-nPR increases from 114.5% to 124.6%; from 10 keV to 30 eV, K-electron-nPR appears to drop by more then 64% from 124.6% to 60%.

Our methods utilizing escape peaks and K-dip spectroscopy have the advantage that the non-proportionality curve can be extended to lower energies than possible with other methods. CCT becomes too inaccurate below 3 keV. With K-dip spectroscopy the curves are extended down to 30 eV. Detailed study of the non-proportionality in the photopeak-nPR just above the K-edge using energy steps as small as 25 eV enables this.

The CCT method has an advantage over K-dip spectroscopy. In K-dip spectroscopy we suppose that in the K-cascade a set of low energy electrons are emitted from the atom and each produces an ionization track. We assumed that these tracks do not interact with the track created by the K-shell photoelectron. In that case the K-dip spectroscopy method provides us like the CCT method the genuine electron response. However, when tracks do influence each other, i.e., when the number of photons produced by the photoelectron is affected by the tracks from the cascade products, an error is introduced In this regard CCT may have an intrinsic advantage over the K-dip spectroscopy, by exciting the crystal with essentially just one electron at a time.

**Acknowledgments**

The research leading to these results has received funding from the Netherlands Technology Foundation (STW), Saint Gobain, crystals and detectors division, Nemours, France, and by the European Community's Seventh Framework Programme (FP7/2007-2013) under grant agreement n° 226716. We thank the scientists and technicians of the X-1 beamline at the Hamburger Synhrotronstrahlungslabor (HASY-LAB) synchrotron radiation facilities for their assistance.

# References

1. R. Hofstadter, Phys. Rev. vol. 74, p. 100, (1948).
2. E.V.D. van Loef, P. Dorenbos, C.W.E. van Eijk, K. Kramer, H.U. Gudel, Appl. Phys. Lett., vol. 79, iss. 10, pp. 1573-1575, (2001).
3. N.J. Cherepy, S.A. Payne, S.J. Asztalos, G. Hull, J.D. Kuntz, T. Niedermayr, S. Pimputkar, J.J. Roberts, R.D. Sanner, T.M. Tillotson, E. van Loef, C.M. Wilson, K.S. Shah, U.N. Roy, R. Hawrami, A. Burger, L.A. Boatner, W.-S. Choong, W.W. Moses, IEEE Trans. Nucl. Sci., vol. 56, no. 3, pp. 873-880 (2009).
4. D. Engelkemeir, Rev. Sci. Instrum., vol. 27, no. 8, pp. 589-591 (1956).
5. C. D. Zebry, A. Meyer, R.B. Murray, Nucl. Instr. Meth., vol. 12, pp. 115-123 (1961).
6. R. B. Murray, A. Meyer, Phys. Rev., vol. 122, pp. 815-826 (1961).
7. P. A. Rodnyi, P. Dorenbos, C. W. E. van Eijk, Phys. Stat. Sol. (b) vol. 187, pp. 15-29 (1995).
8. P. Dorenbos, J. T. M. De Haas, and C. W. E. Van Eijk, IEEE Trans. Nucl. Sci., vol. 42, no. 6, pp. 2190-2202 (1995).
9. J. D. Valentine, B. D. Rooney, J. Li, IEEE Trans. Nucl. Sci., vol. 45, no. 3, pp. 512-517 (1998).
10. K.D. Ianakiev, M.E. Abhold, B.S. Alexandrov, M.C. Browne, R.M. Williams, P.B. Littlewood, Nucl. Instr. Meth. Phys. Res. A 579, pp. 34–37 (2007).
11. M. Moszynski, A. Nassalski, A. Syntfeld-Kazuch, L. Swiderski, T. Szczesniak, IEEE Trans. Nucl. Sci., vol. 55, no. 3, pp. 1062-1068 (2008).
12. G. Bizarri, W. W. Moses, J. Singh, A. N. Vasil'ev, R. T. Williams, J. Appl. Phys., vol. 105, 044507 (2009).
13. P.A. Rodnyi, *Physical Processes in Inorganic Scintillators*, CRC Press (1997).
14. P. A. Rodnyi, Rad. Measur. Vol. 29, No. 3-4, pp. 235-242 (1998).
15. B. D. Rooney, J. D. Valentine, IEEE Trans. Nucl. Sci., vol. 43, no. 3, pp. 1271-1276 (1996).
16. W.-S. Choong, K. M. Vetter, W. W. Moses, G. Hull, S. A. Payne, N. J. Cherepy, J. D. Valentine, IEEE Trans. Nucl. Sci., vol. 55, no. 3, pp. 1753-1758 (2008).
17. A. J. L. Collinson, R. Hill, Proc. Phys. Soc., vol. 81, pp. 883-892 (1963).
18. L.R. Wayne, W.A. Heindl, P.L. Hink, R.E. Rothschild, Nucl. Instr. and Meth. A 411 pp. 351-364 (1998).
19. I.V. Khodyuk, J.T.M. de Haas, P. Dorenbos, submitted to IEEE Trans. Nucl. Sci.
20. J. T. M. de Haas, P. Dorenbos, C. W. E. van Eijk, Nucl. Instr. Meth. Phys. Res. A, vol. 537, pp. 97-100 (2005).
21. J. T. M. de Haas, P. Dorenbos, IEEE Trans. Nucl. Sci. vol. 55, no. 3, pp. 1086-1092 (2008).
22. M. Moszyn´ski, M. Balcerzyk, W. Czarnacki, M. Kapusta, W. Klamra, A. Syntfeld, M. Szawlowski, IEEE Trans. Nucl. Sci., vol. 51, no. 3, (2004).
23. E. V. D. van Loef, W. Mengesha, J. D. Valentine, P. Dorenbos, C. W. E. van Eijk, IEEE Trans. Nucl. Sci., vol. 50, no. 1, pp. 155-158 (2003).
24. G.C. Meggitt, Nucl. Instr. And Meth. 83 pp.313 (1970).
25. B. D. Rooney, J. D. Valentine, IEEE Trans. Nucl. Sci., vol. 44, no. 3 (1997).
26. G. Hull, W.-S. Choong, W.W. Moses, G. Bizarri, J.D. Valentine, S.A. Payne, N.J. Cherepy, B.W. Reutter, IEEE Trans. Nucl. Sc. vol. 56, no. 1 (2009).

Table 1. Properties of Iodine X-ray fluorescence transitions. The type of transition (Line), the subshell and orbital where it originates from, its energy (in keV), and probability are given.

| Line | Subshell | Orbital | Energy | Probability |
|---|---|---|---|---|
| $K_{\alpha 1}$ | $L_3$ | $2p_{3/2}$ | 28.612 | 0.5338 |
| $K_{\alpha 2}$ | $L_2$ | $2p_{1/2}$ | 28.317 | 0.2875 |
| $K_{\beta 1}$ | $M_3$ | $3p_{3/2}$ | 32.294 | 0.0947 |
| $K_{\beta 2}$ | $N_{2,3}$ | $4p_{1/2}, 4p_{3/2}$ | 33.046 | 0.0326 |
| $K_{\beta 3}$ | $M_2$ | $3p_{1/2}$ | 32.238 | 0.0491 |

Fig. 1. X-1 beamline experimental set-up at the Hamburger Synhrotronstrahlungslabor (HASYLAB) synchrotron radiation facility in Hamburg, Germany.

Fig. 2. Pulse height spectrum measured with NaI:Tl at 40 keV monochromatic X-ray irradiation. a – photopeak, b – escape peaks. The inset shows the escape peaks on an expanded scale. The solid line in the inset is the result of a fit with five Gaussian peak.

Fig. 3. Most probable K X-ray fluorescence transitions in iodine atomic shells.

Fig. 4. The scintillation yield of NaI:Tl measured with a Hamamatsu R6231-100 PMT versus X-ray energy. The right scale shows the peak position of the photopeak and the left scale corresponding number of the photoelectrons $N_{phe}^{PMT}$. Inset: expanded scale at energies near the Iodine K-electron binding energy.

Fig. 5. Photopeak non-proportional response of NaI:Tl as a function of X-ray energy at 5 keV intervals.

Fig. 6. Photopeak non-proportional response of NaI:Tl as a function of X-ray energy near the Iodine K-electron binding energy at 25 eV intervals.

Fig. 7. Energy resolution of the X-ray photopeak recorded with the NaI:Tl scintillator as a function of X-ray energy.

Fig. 8. Energy resolution of NaI:Tl as function of the number of photoelectrons $N_{phe}^{PMT}$. Solid line – contribution due to Poisson statistics. The inset shows on an expanded scale the resolution near the Iodine K-electron binding energy.

Fig. 9. Photon non-proportional response of NaI:Tl as a function of deposited energy. Black solid circles, photopeak-nPR; blue open squares, $K_\alpha$ escape-nPR; red open circles, $K_\beta$ escape-nPR . The solid curve shows the X-ray attenuation length for NaI.

Fig. 10. Escape non-proportional response of NaI:Tl as a function of deposited energy. Black open squares, $K_\alpha$ escape-nPR, red solid circles, $K_\beta$ escape-nPR.The arrows indicate the locations of K, L, and M-shell electron binding energies of Iodine and Sodium.

Fig. 11. Comparison of electron non-proportional response as a function of photoelectron energy inferred using K-dip spectroscopy with other data.

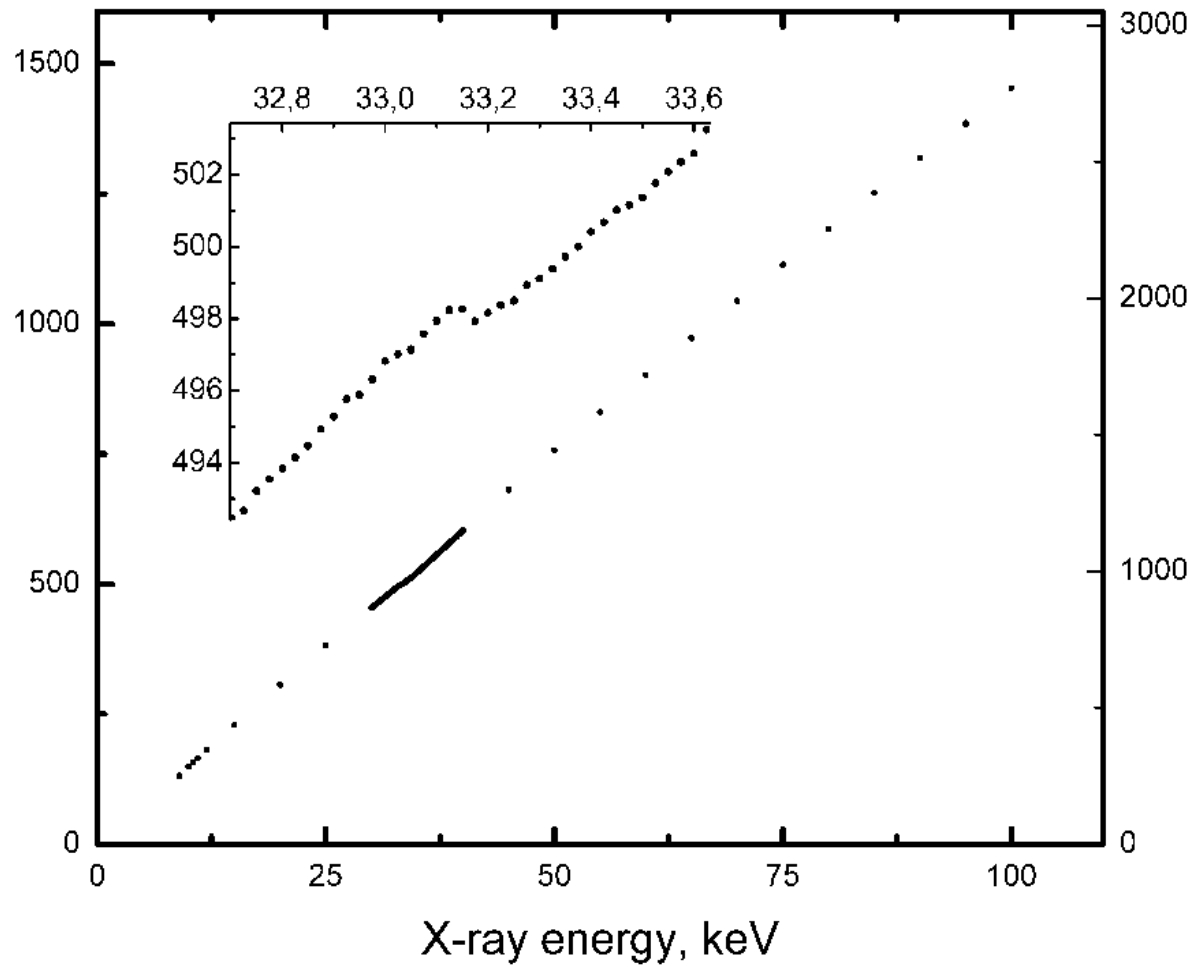

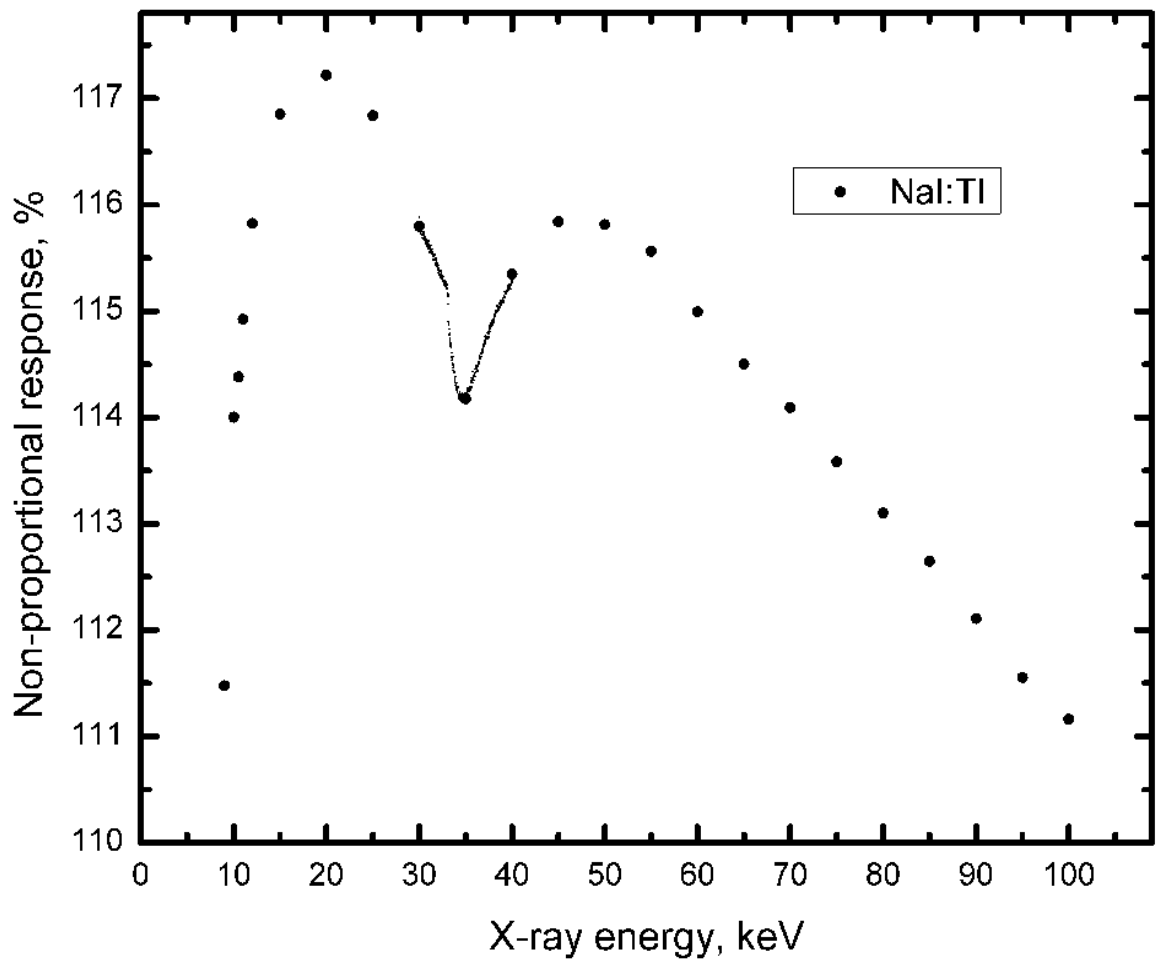

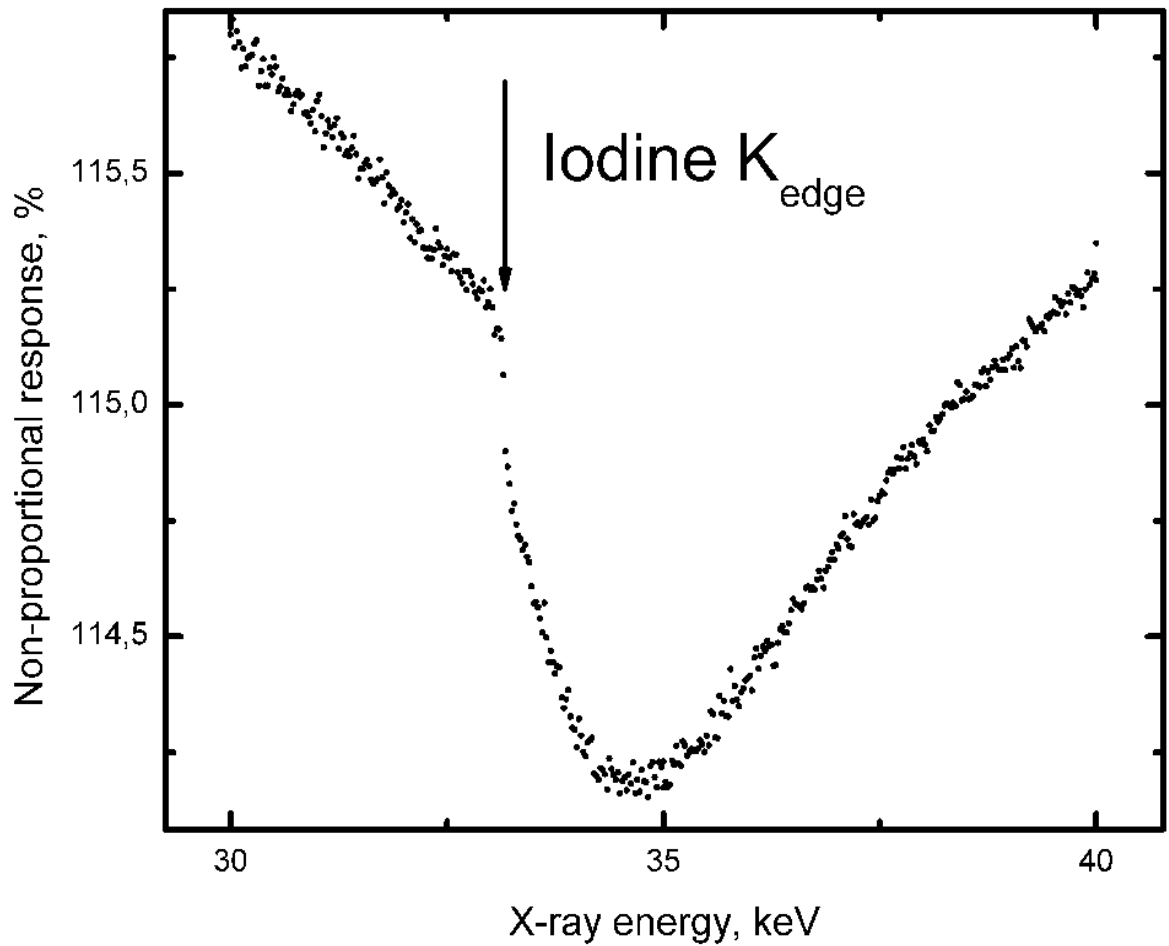

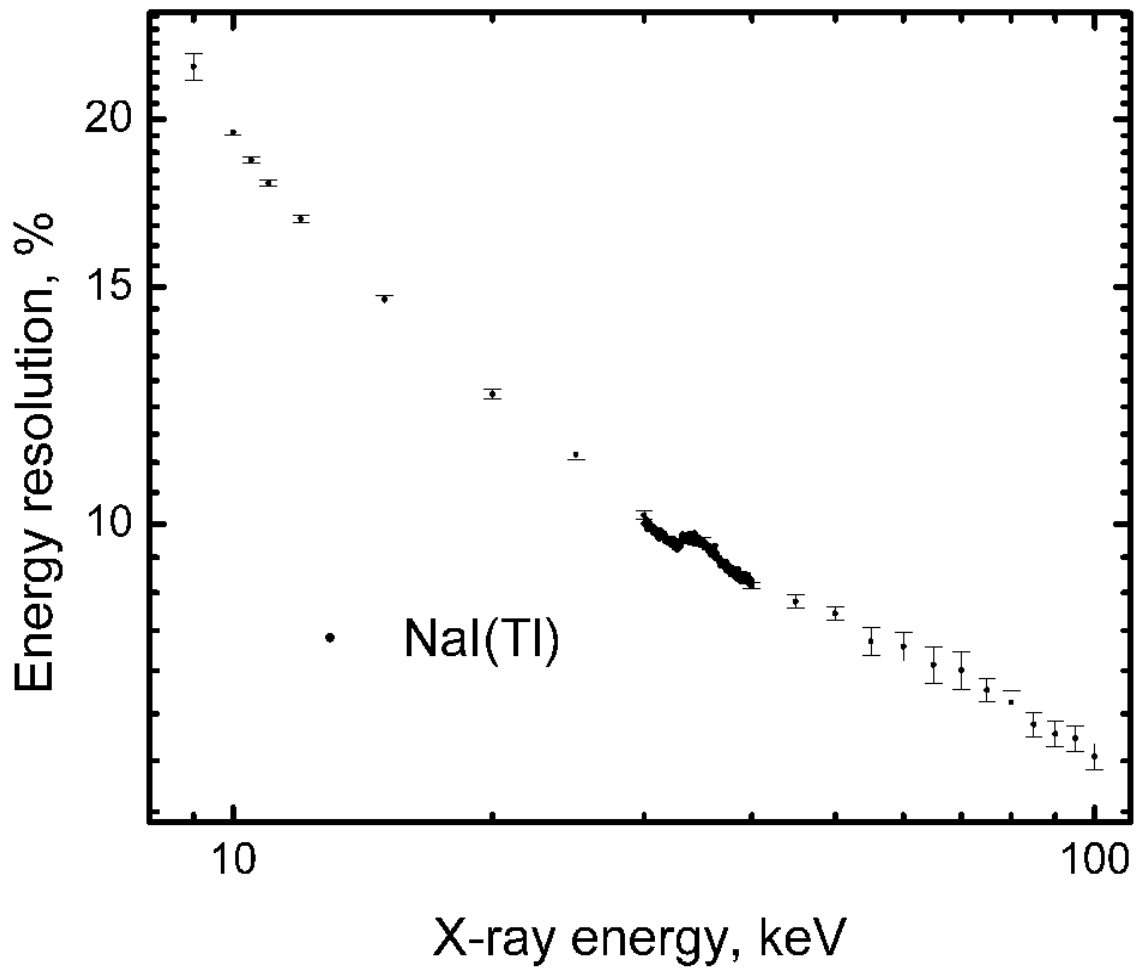

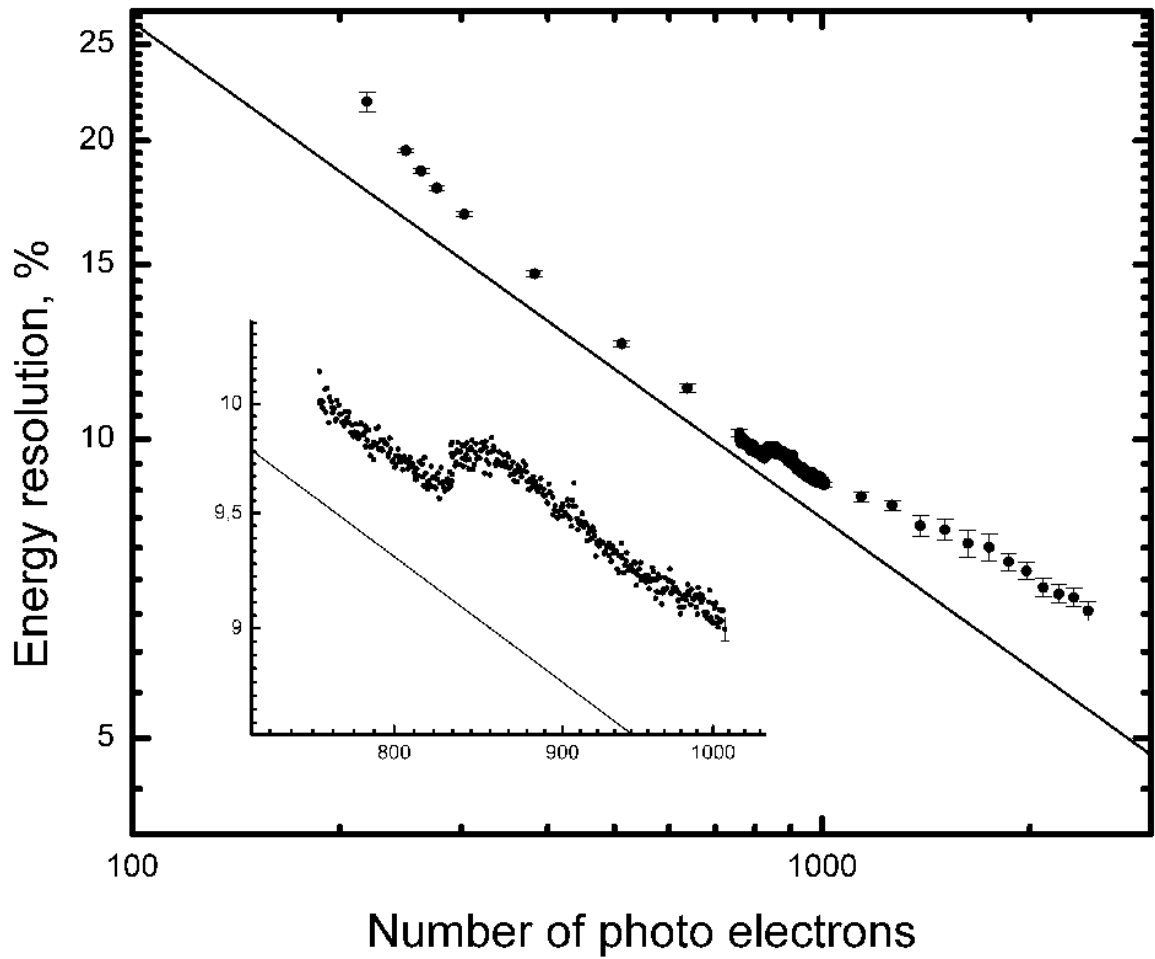

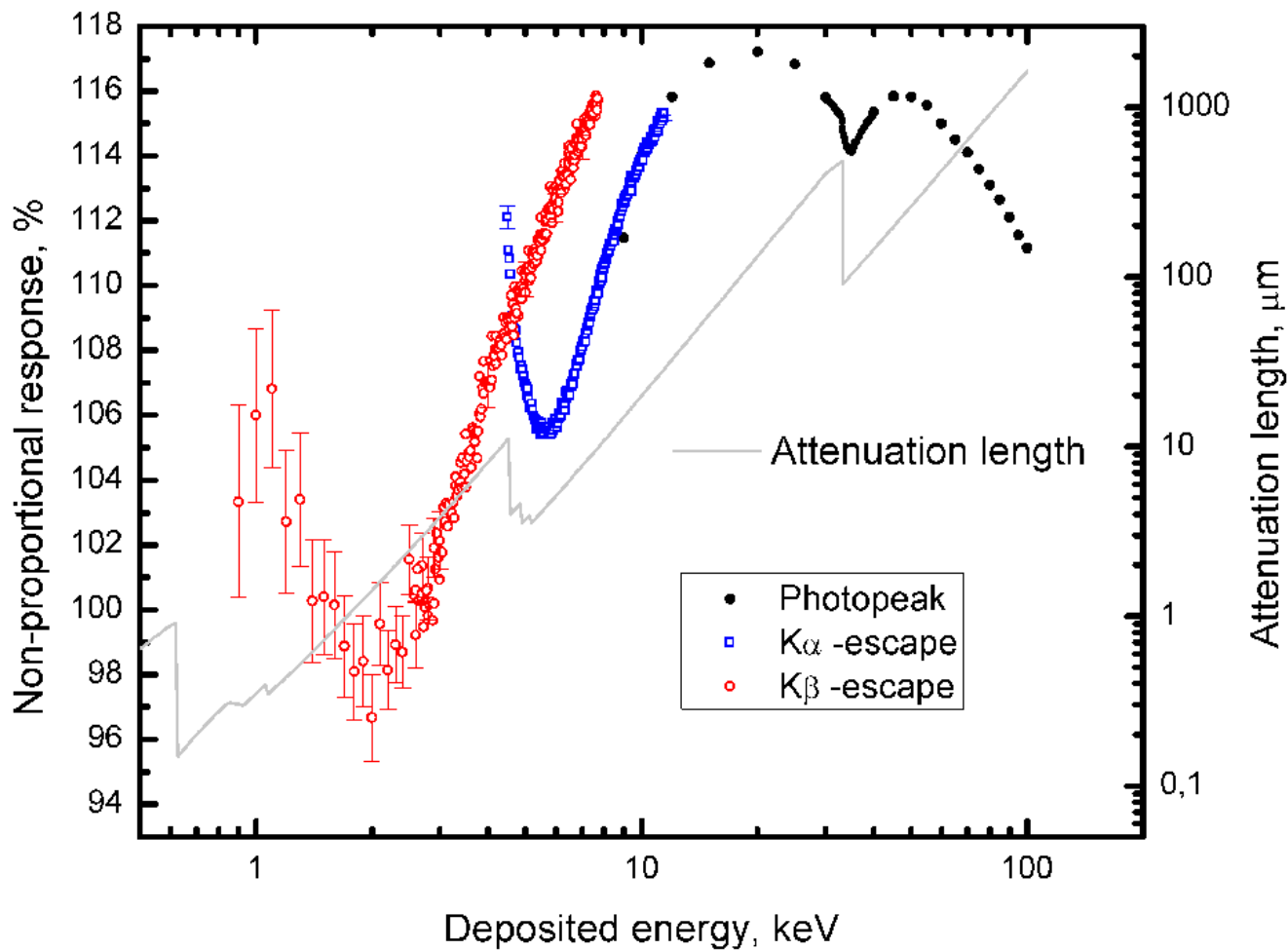

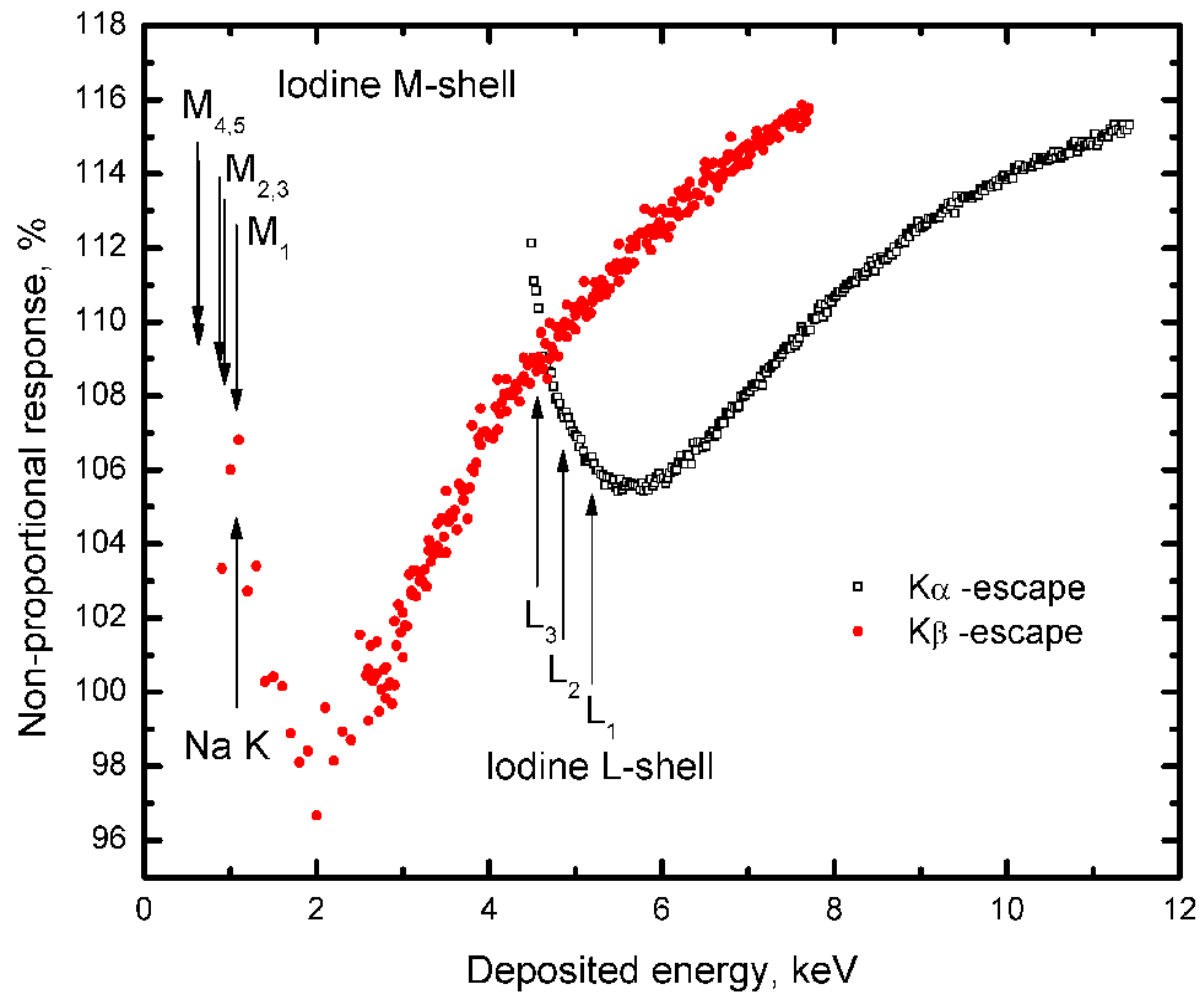

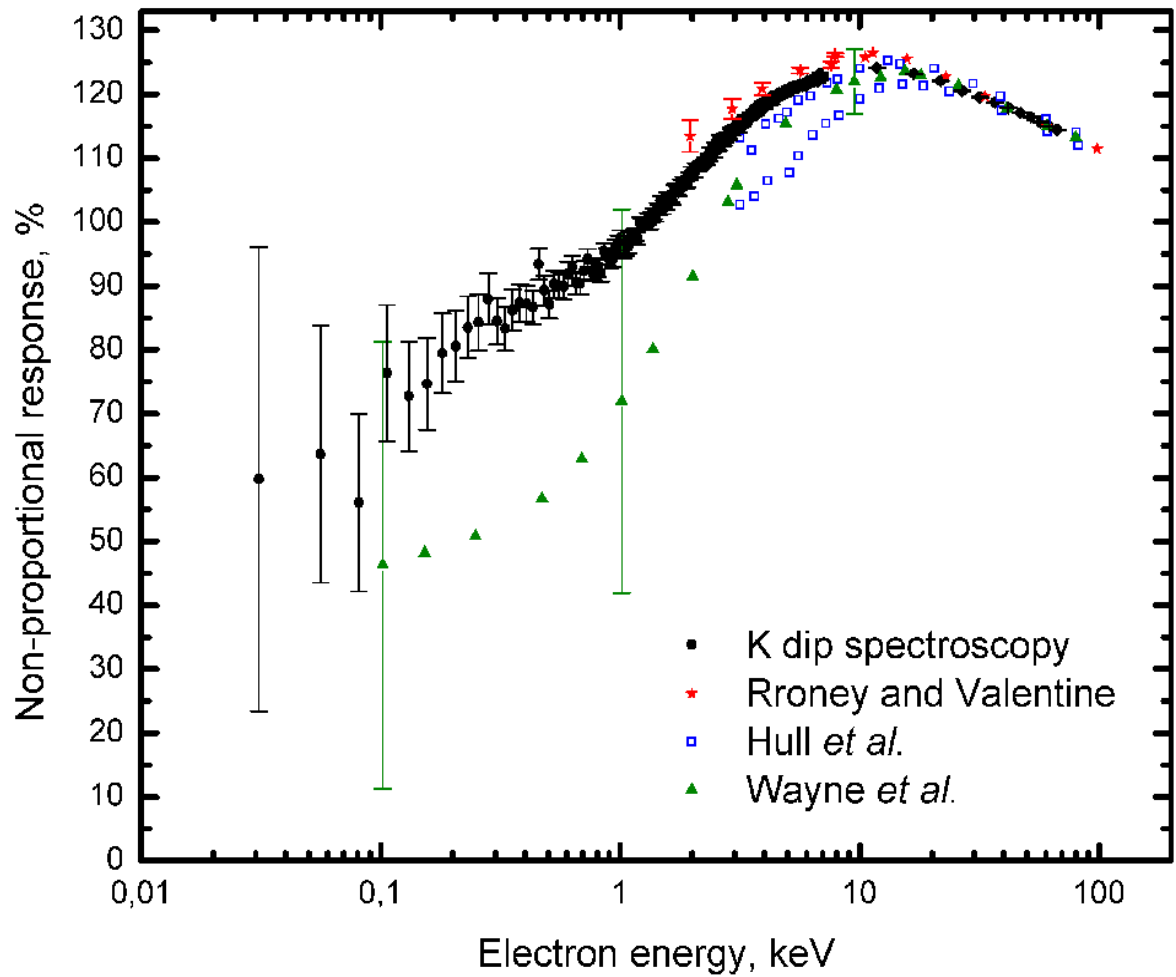

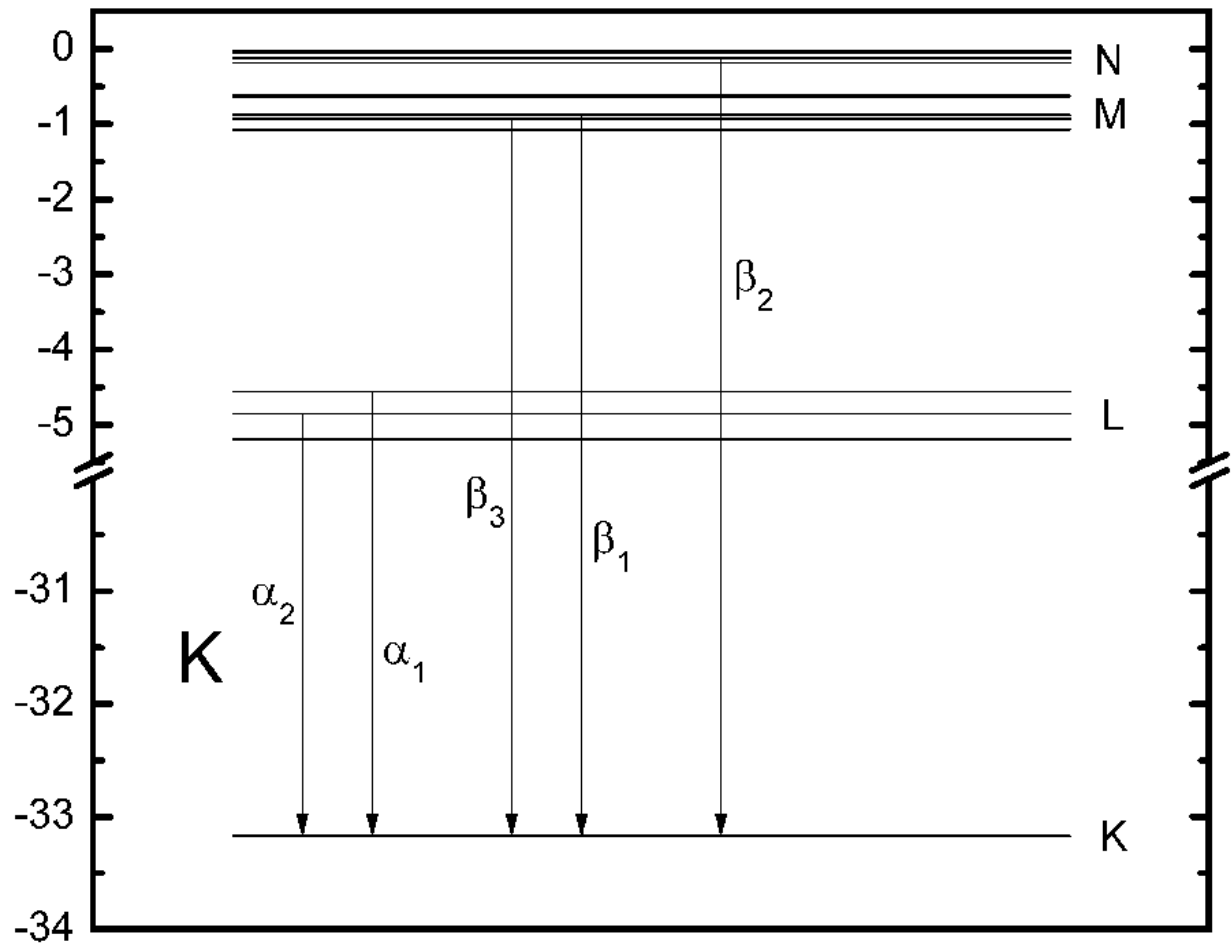

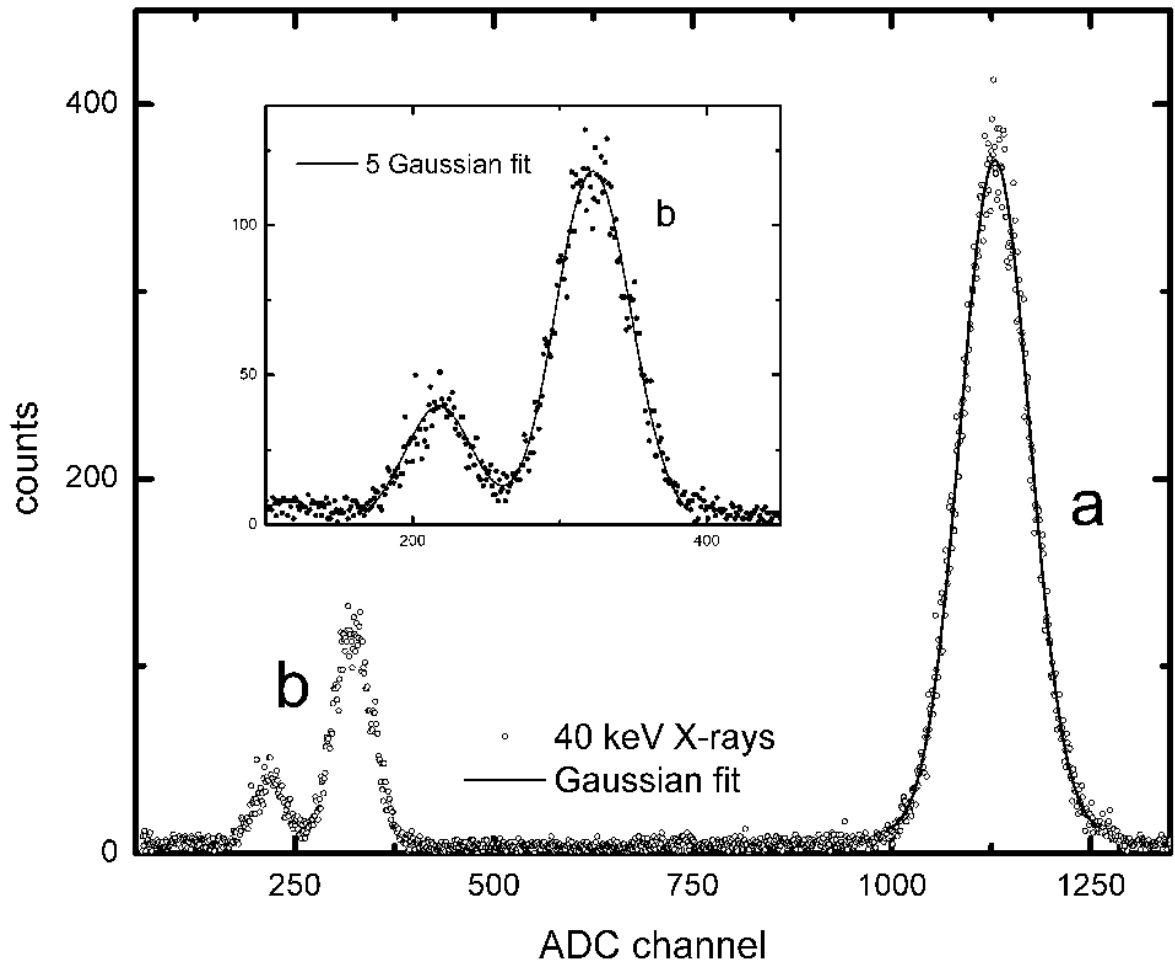

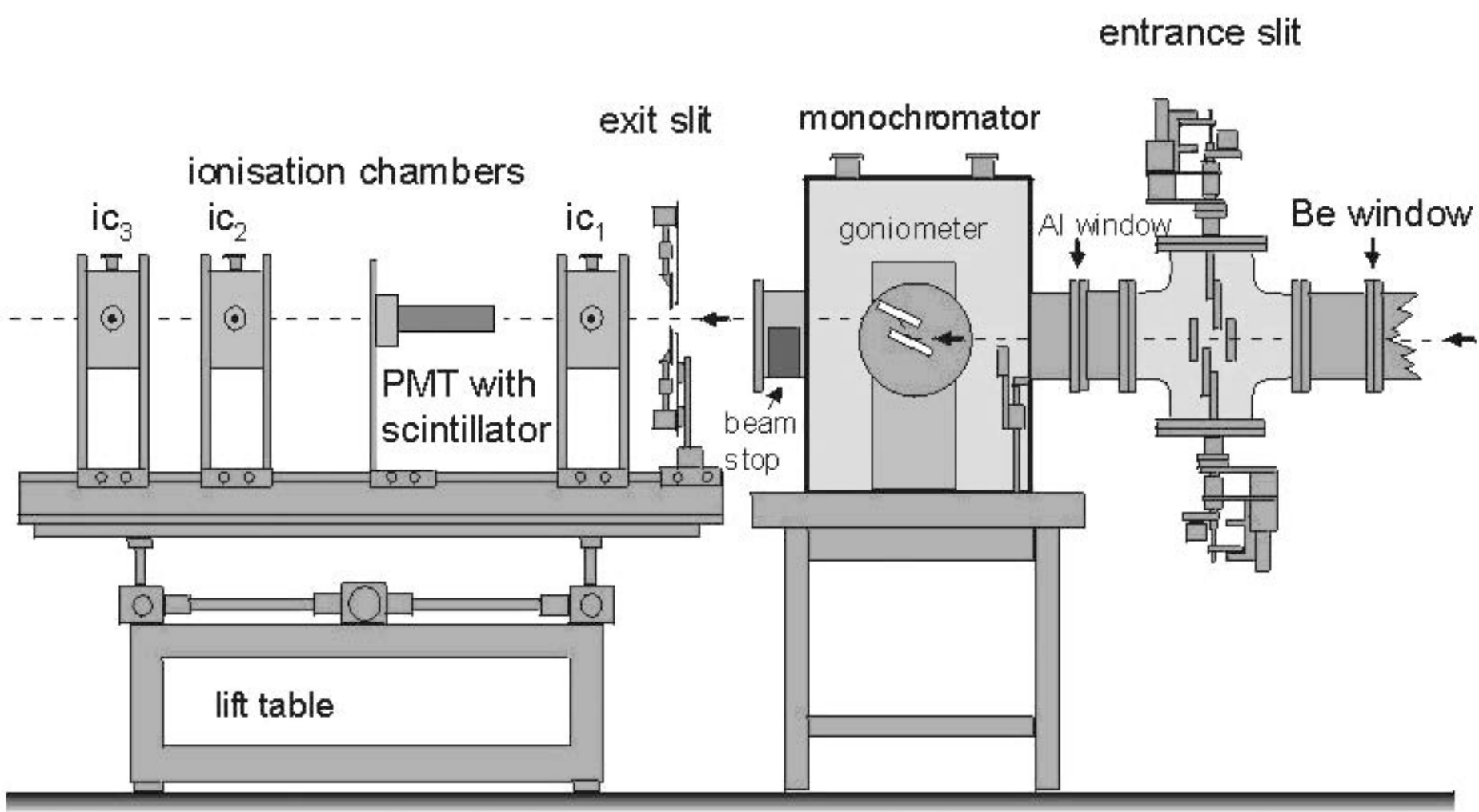